\documentclass[journal]{IAENGtran}
\ifCLASSINFOpdf
   \usepackage[pdftex]{graphicx}
   \DeclareGraphicsExtensions{.pdf,.jpeg,.png}
\else
   \usepackage[dvips]{graphicx}
   \DeclareGraphicsExtensions{.eps}
\fi

\begin{document}
%
\title{Decision Making via AHP}
%
%
%

\author{M. Andrecut 
 \thanks{January 8, 2014. M. Andrecut is with Unlimited Analytics Inc., 
 Calgary, Alberta, T3G 5Y8, Canada, Email: mircea.andrecut@gmail.com 
 }
}

\maketitle

\pagestyle{empty}
\thispagestyle{empty}

\begin{abstract}
The Analytic Hierarchy Process (AHP) is a procedure for establishing
priorities in multi-criteria decision making problems. Here we discuss 
the Logarithmic Least Squares  (LLS) method for the AHP and group-AHP, which provides an exact and unique  solution
for the priority vector. Also, we show that for the group-AHP, the
LLS method is equivalent with the minimization of the weighted sum
of generalized Kullback-Leibler divergences, between the group-priority
vector and the priority vector of each expert.
\end{abstract}

\begin{IAENGkeywords}
decision making, analytic hierarchy process, Kullback-Leibler divergence.
\end{IAENGkeywords}

%
\IAENGpeerreviewmaketitle

\section{Introduction}
%
%
%
%
\IAENGPARstart{T}{he} Analytic Hierarchy Process (AHP) is a popular multi-criteria decision
making method, with important psychometric, economic, industrial and
military applications \cite{key-1,key-2}. The main role of the AHP
is to provide solutions to decision problems, where several alternatives
for obtaining given objectives are compared under different criteria.
The AHP represents the objectives, the alternatives and the criteria
of the problem in a hierarchical structure. The decision maker (expert)
produces a series of pairwise comparison judgments of the relative
strength of the alternatives at the same level of the hierarchy. These
judgments are then converted into numbers using a ratio scale, and
organized in judgment matrices, which are then used to establish the
decision weights, or the priorities for alternatives. The most familiar
method to estimate the priorities from a judgment matrix is the Saaty's
Eigenvector (SE) method \cite{key-1,key-2}, which is based on the
principal eigenvector of the judgment matrix. However, the SE method
has been criticized both from prioritization and consistency points
of view, and several new techniques based on minimization methods
have been proposed, such as: Least Squares (LS) \cite{key-3}, Logarithmic
Least Squares (LLS) \cite{key-4,key-5,key-6}, Weighted Least Squares
(WLS) \cite{key-7}, Logarithmic Least Absolute Values (LLAV) \cite{key-8},
and Singular Value Decomposition (SVD) \cite{key-9}. With the exception
of the LLS method, the other minimization methods are difficult to
apply, and can even result in several minima which makes the choice
ambiguous \cite{key-9}. Therefore, the most important alternative
to the SE method is the LLS approach, which provides an unique solution,
by minimizing a logarithmic objective function, subject to a multiplicative
constraint between the components of the priority vector. Here, we discuss the LLS method for the AHP and group-AHP, which provides an exact and unique solution
for the priority vector.
Also, we show that for the group-AHP, the LLS approach is equivalent
with the minimization of the weighted sum of generalized Kullback-Leibler
divergences, between the group-priority vector and the priority vector
of each expert.

\section{AHP}

Let us briefly define the AHP and its essential elements \cite{key-1,key-2}.
We assume that there are $N$ alternatives $A_{n}$, $n=1,2,...,N$,
and an expert provides his opinions on each pair of them $(A_{i},A_{j})$,
expressing the strength $a_{ij}$ of one factor $A_{i}$ over the
second one $A_{j}$, using a numerical ratio scale. The comparison scale ranges
from $a_{ij}=1/9$ for $A_{i}$ least preferred than $A_{j}$, to
$a_{ij}=1$ for $A_{i}$ equally preferred to $A_{j}$, and to $a_{ij}=9$
for $A_{i}$ extremely preferred to $A_{j}$, covering the entire
spectrum of the comparison. Thus, the expert creates a $N\times N$
reciprocal judgment matrix $A=[a_{ij}]_{N\times N}$, where: \begin{equation}
a_{ij}=\frac{1}{a_{ji}},\; a_{ij}>0,\; i,j=1,2,\ldots,N,\end{equation}
such that the elements of the main diagonal are all equal to $1$,
and the symmetrical elements are mutually reciprocal. Therefore, only $N(N-1)/2$
judgments are required to construct the matrix. The goal of the AHP
is to obtain a vector of priorities $w=[w_{1},w_{2},\ldots,w_{N}]^{T}$,
with the normalized components:\begin{equation}
\sum_{i=1}^{N}w_{i}=1,\; w_{i}>0,\; i=1,2,\ldots,N,\end{equation}
such that the $N\times N$ matrix of ratios $W=[w_{i}/w_{j}]_{N\times N}$,
approximates the judgment matrix $A$, i.e. $W\simeq A$. 

A judgment matrix $A=[a_{ij}]_{N\times N}$, satisfying:\begin{equation}
a_{ij}=\frac{u_{i}}{u_{j}},\; u_{i}>0,\; i,j=1,2,\ldots,N,\end{equation}
is said to be consistent. It follows immediately that a consistent
judgment matrix also satisfies the transitivity relation:\begin{equation}
a_{ik}=a_{ij}a_{jk},\; i,j,k=1,2,\ldots,N.\end{equation}
It is also easy to see that every consistent matrix is also a judgment
matrix. Also, if $A$ is consistent, then every element of $A$ can
be determined from the first row of $A$, since:\begin{equation}
a_{ik}=\frac{a_{1k}}{a_{1i}},\; i,k=1,2,\ldots,N,\end{equation}
and therefore $A$ is a rank one matrix with exactly one non-zero
eigenvalue. Moreover, if $A$ is consistent, we have:
\begin{equation}
Au=Nu,\end{equation}
and the single non-zero value of $A$ is $\lambda=N$. 

The normalized
eigenvector $w=[w_{1},w_{2},\ldots,w_{N}]^{T}$, with the components:\begin{equation}
w_{i}=\frac{u_{i}}{\sum_{j=1}^{N}u_{j}},\; j=1,2,\ldots,N,\end{equation}
is the vector of priorities derived from $A$. 

In general, the judgments $a_{ij}$ are rarely perfect, and the transitivity
relation is therefore frequently violated. In this case, the judgment
matrix is said to be inconsistent. The degree of inconsistency varies
from subjective or objective reasons, and in general rises with the
size of the matrix. In this case, there is not an unique way of deriving
the priority vector $w$, and various methods may produce different
results. If the judgment matrix $A$ is inconsistent then, using the
Perron-Frobenius theorem for positive matrices, one can show that
the largest eigenvalue is $\lambda>N$, and the normalized difference
$\mu=(\lambda-N)/(N-1)$ has been proposed as a measure of the inconsistency
\cite{key-1,key-2}. If the inconsistency $\mu$ is larger than a
given threshold, the AHP recommends to correct the matrix, until near
consistency is reached \cite{key-1,key-2}. Thus, an inconsistent
judgment matrix can be seen as a perturbation of a consistent one.
When the perturbations are small, the maximal eigenvalue is close
to $N$, and the corresponding principal eigenvector is close to the
eigenvector of the unperturbed consistent matrix. A good estimate
of the principal eigenvector can be obtained using the SE method,
i.e. raising the matrix at large powers, normalizing the row sums
each time, and stopping the procedure when the difference between
normalized sums in two consecutive calculations is smaller than a
prescribed value. It has been shown that for small deviations around
the consistent ratios $u_{i}/u_{j}$, the SE method gives reasonably
good approximation of the priority vector. However, when the inconsistencies
are large, it is generally accepted that the SE solution is not satisfactory,
and other methods like the LLS are more appropriate.

\section{LLS method}

The LLS method assumes that the elements of the inconsistent judgment
matrix $A$ are approximated by \cite{key-4,key-5,key-6}: \begin{equation}
a_{ij}=\frac{u_{i}}{u_{j}}\varepsilon_{ij},\; u_{i}>0,\;\varepsilon_{ii}=1,\;\varepsilon_{ij}=1/\varepsilon_{ji},\;\varepsilon_{ij}>0,\end{equation}
where $\varepsilon_{ij}$ are the perturbations from consistency.
A small perturbation means that $\varepsilon_{ij}\simeq1$. Thus,
the approximation matrix is obviously reciprocally symmetric, and
has unit diagonal terms, however it may violate the consistency condition
$a_{ik}=a_{ij}a_{jk}$, due to the included perturbation factor $E=[\varepsilon_{ij}]_{N\times N}$.
The LLS approach is based on the minimization of the sum of squares:\begin{equation}
S=\sum_{i=1}^{N}\sum_{j=1}^{N}\delta_{ij}^{2},\end{equation}
where $\delta_{ij}$ are the errors of the log approximation equations:\begin{equation}
\delta_{ij}=\ln\varepsilon_{ij}=\ln a_{ij}-\ln u_{i}+\ln u_{j},\; i,j=1,2,\ldots,N.\end{equation}
It has been shown \cite{key-4,key-5,key-6} that when $\delta_{ij}$
are independent and normally distributed, with zero mean and common
variance $\sigma^{2}$, the solution $u=[u_{1},u_{2},\ldots,u_{N}]^{T}$
for the problem:\begin{equation}
\min_{u_{1},u_{2},\ldots,u_{N}}\sum_{i=1}^{N}\sum_{j=1}^{N}[\ln a_{ij}-\ln u_{i}+\ln u_{j}]^{2},\end{equation}
subject to the constraint:\begin{equation}
\prod_{i=1}^{N}u_{i}=1,\end{equation}
is unique and it is given by the geometric means of the rows of the
matrix $A$:\begin{equation}
u_{i}=\prod_{j=1}^{N}a_{ij}^{1/N},\; i=1,2,\ldots,N.\end{equation}
Thus, the vector of priorities $w$ is obtained by normalizing the
components of $u$, using (7), such that $\sum_{i=1}^{N}w_{i}=1$.
Also, it has been shown that the LLS approach leads to an unbiased
estimation of $\sigma^{2}$ (the variance of perturbations) as a measure
of consistency \cite{key-4}:\begin{equation}
\sigma^{2}=\frac{1}{(N-1)(N-2)}\sum_{i=1}^{N}\sum_{j=1}^{N}[\ln a_{ij}-\ln u_{i}+\ln u_{j}]^{2}.\end{equation}

One obtains an exact and unique solution of this problem, by directly
solving the minimization problem:\begin{equation}
\min_{u_{1},u_{2},\ldots,u_{N}}S(u_{1},u_{2},\ldots,u_{N}),\end{equation}
where
\begin{equation}
S(u_{1},u_{2},\ldots,u_{N})=\sum_{i=1}^{N}\sum_{j=1}^{N}[\ln a_{ij}-\ln u_{i}+\ln u_{j}]^{2}.
\end{equation}
The error function $S$ can be rewritten as:\begin{equation}
S(x_{1},x_{2},\ldots,x_{N})=\sum_{i=1}^{N}\sum_{j=1}^{N}[b_{ij}-x_{i}+x_{j}]^{2},\end{equation}
where $b_{ij}=\ln a_{ij}$ and $x_{i}=\ln u_{i}$. 

Obviously $S(x_{1},x_{2},\ldots,x_{N})$
is convex and has an unique minimum at $(x_{1},x_{2},\ldots,x_{N})$
where:\begin{equation}
\frac{\partial}{\partial x_{k}}S(x_{1},x_{2},\ldots,x_{N})=0,\; k=1,2,\ldots,N,\end{equation}
which is equivalent with the system of linear equations: \begin{equation}
x_{k}-\frac{1}{N}\sum_{i=1}^{N}x_{i}=\left\langle B\right\rangle _{k},\; k=1,2,\ldots,N,\end{equation}
where\begin{equation}
\left\langle B\right\rangle _{k}=\frac{1}{N}\sum_{j=1}^{N}b_{kj},\end{equation}
is the average of the row $k$ of the matrix $B=[b_{ij}]_{N\times N}=[\ln a_{ij}]_{N\times N}$.

This system can be written in a matrix form as following:\begin{equation}
Qx=d,\end{equation}
where $Q=[q_{ij}]_{N\times N}$ and $d=[d_{1},d_{2},\ldots,d_{N}]^{T}$
are given by:\begin{eqnarray}
q_{ij} & = & \left\{ \begin{array}{ccc}
1-\frac{1}{N} & if & i=j\\
-\frac{1}{N} & if & i\neq j\end{array}\right.,\; i,j=1,2,\ldots,N,\end{eqnarray}
and respectively:\begin{equation}
d_{i}=\left\langle B\right\rangle _{i},\; i=1,2,\ldots,N.\end{equation}

The minimum norm solution of the linear system $Qx=d$ is given by:
\begin{equation}
x_{k}=\left\langle B\right\rangle _{k}-\frac{1}{N}\sum_{i=1}^{N}\left\langle B\right\rangle _{i},\; k=1,2,\ldots,N.
\end{equation}
We also have:
$$
\frac{1}{N}\sum_{i=1}^{N}\left\langle B\right\rangle _{i}=\frac{1}{N^{2}}\sum_{i=1}^{N}\sum_{j=1}^{N}b_{ij}=
$$
\begin{equation}
=\frac{1}{N^{2}}\sum_{i=1}^{N}\sum_{j=1}^{N}\ln a_{ij}=0,
\end{equation}
since:\begin{equation}
a_{ij}a_{ji}=1\Leftrightarrow\ln a_{ij}+\ln a_{ji}=0.\end{equation}
Therefore, the solution of the linear system is:\begin{equation}
x_{k}=\left\langle B\right\rangle _{k},\; k=1,2,\ldots,N.\end{equation}
From here we obtain:
$$
u_{k}=\exp\left(\left\langle B\right\rangle _{k}\right)=
$$
\begin{equation}
=\exp\left(\frac{1}{N}\sum_{j=1}^{N}\ln a_{kj}\right),\; k=1,2,\ldots,N.
\end{equation}
Also, we observe that:
$$
\exp\left(\frac{1}{N}\sum_{j=1}^{N}\ln a_{kj}\right)=\exp\left(\ln\left(\prod_{j=1}^{N}a_{kj}\right)^{1/N}\right)=
$$
\begin{equation}
=\prod_{j=1}^{N}a_{kj}^{1/N}.
\end{equation}

The above analysis represents the proof of the following theorem and corollary:

{\bf Theorem:} Let $A=[a_{ij}]$ be a  $N\times N$ judgement matrix, and 
$B=[b_{ij}]$ a consistent $N\times N$ matrix, with $b_{ij}=u_{i}/u_{j}$. Then, 
the best consistent matrix approximation of $A$ is obtained for the vector $u=[u_{1},...,u_{N}]^{T}$, with the components 
equal to the geometric mean of the corresponding rows of $A$, i.e.:
\begin{equation}
u_{i}=\prod_{j=1}^{N}a_{ij}^{1/N}.
\end{equation}

{\bf Corollary:} 
If $A=[a_{ij}]$ is a  $N\times N$ judgement matrix, and 
$B=[b_{ij}]=[u_{i}/u_{j}]$ is a consistent $N\times N$ matrix, with:
$
u_{i}=\prod_{j=1}^{N}a_{ij}^{1/N},
$
then:
\begin{equation}
d(A,B)=\sqrt{\sum_{i=1}^{N}\sum_{j=1}^{N}[\ln a_{ij}-\ln b_{ij}]^{2}}.
\end{equation}
is the minimal distance from $A$ to any consistent matrix.

Thus, the components of the priority vector $w$ are obtained by normalizing
the components of the vector $u$:\begin{equation}
w_{k}=\frac{\prod_{j=1}^{N}a_{kj}^{1/N}}{\sum_{i=1}^{N}\prod_{j=1}^{N}a_{ij}^{1/N}},\; k=1,2,\ldots,N,\end{equation}
such that $\sum_{k=1}^{N}w_{k}=1$. 

\section{LLS method for group-AHP}

We consider $M$ experts, judging $N$ alternatives $A_{n}$, $n=1,2,...,N$.
Each expert is characterized by a different weight $\alpha_{m}>0$,
$m=1,2,\ldots M$, corresponding to the expert's level, such that:
\begin{equation}
\sum_{m=1}^{M}\alpha_{m}=1.\end{equation}
Also, each expert produces a $N\times N$ reciprocal judgment matrix
$A^{(m)}=[a_{ij}^{(m)}]_{N\times N}$, where: \begin{equation}
a_{ij}^{(m)}=\frac{1}{a_{ji}^{(m)}},\; a_{ij}^{(m)}>0.\end{equation}
The statistical model for each expert is obtained by introducing the
multiplicative errors $\varepsilon_{ij}^{(m)}$, such that:

\begin{equation}
a_{ij}^{(m)}=\frac{u_{i}}{u_{j}}\varepsilon_{ij}^{(m)},\end{equation}
where\begin{equation}
u_{i}>0,\;\varepsilon_{ii}^{(m)}=1,\;\varepsilon_{ij}^{(m)}=1/\varepsilon_{ji}^{(m)},\;\varepsilon_{ij}^{(m)}>0.\end{equation}
Our goal is to find the estimates $u_{i}$, using the LLS approach
discussed in the previous section. Therefore, we consider the unconstrained
minimization problem:\begin{equation}
\min_{u_{1},u_{2},\ldots,u_{N}}S(u_{1},u_{2},\ldots,u_{N}),\end{equation}
where
$$
S(u_{1},u_{2},\ldots,u_{N})=
$$
\begin{equation}
=\sum_{m=1}^{M}\sum_{i=1}^{N}\sum_{j=1}^{N}\alpha_{m}\left[\ln a_{ij}^{(m)}-\ln u_{i}+\ln u_{j}\right]^{2},
\end{equation}
is the weighted sum of approximation errors. The problem can be rewritten
as:\begin{equation}
S(x_{1},x_{2},\ldots,x_{N})=\sum_{m=1}^{M}\sum_{i=1}^{N}\sum_{j=1}^{N}\alpha_{m}[b_{ij}^{(m)}-x_{i}+x_{j}]^{2},\end{equation}
where $b_{ij}^{(m)}=\ln a_{ij}^{(m)}$ and $x_{i}=\ln u_{i}$. The
minimum condition:\begin{equation}
\frac{\partial}{\partial x_{k}}S(x_{1},x_{2},\ldots,x_{N})=0,\; k=1,2,\ldots,N,\end{equation}
is equivalent with the system of linear equations: \begin{equation}
x_{k}-\frac{1}{N}\sum_{i=1}^{N}x_{i}=\left\langle B\right\rangle _{\alpha,k},\; k=1,2,\ldots,N,\end{equation}
where\begin{equation}
\left\langle B\right\rangle _{\alpha,k}=\frac{1}{N}\sum_{m=1}^{M}\sum_{j=1}^{N}\alpha_{m}b_{kj}^{(m)},\end{equation}
is the weighted average of the row $k$ of all logarithmic matrices $B^{(m)}=[b_{ij}^{(m)}]_{N\times N}$.
Again, this is a linear system of the form $Qx=d$, with  $d_{k}=\left\langle B\right\rangle _{\alpha,k}$,
and therefore the solution is:\begin{equation}
x_{k}=\left\langle B\right\rangle _{\alpha,k},\; k=1,2,\ldots,N,\end{equation}
and respectively:\begin{equation}
u_{k}=\exp\left(\left\langle B\right\rangle _{\alpha,k}\right),\; k=1,2,\ldots,N.\end{equation}
Also, we observe that:
$$
\exp\left(\left\langle B\right\rangle _{\alpha,k}\right)=\exp\left(\frac{1}{N}\sum_{m=1}^{M}\sum_{j=1}^{N}\alpha_{m}\ln a_{ij}^{(m)}\right)=
$$
$$
=\exp\left(\ln\left(\prod_{m=1}^{M}\prod_{j=1}^{N}\left(a_{kj}^{(m)}\right)^{\alpha_{m}/N}\right)\right)=
$$
\begin{equation}
=\prod_{m=1}^{M}\prod_{j=1}^{N}\left(a_{kj}^{(m)}\right)^{\alpha_{m}/N}.
\end{equation}
Thus, the components of the group-priority vector $w$
are obtained by normalizing the components of the vector $u$:
\begin{equation}
w_{k}=\frac{\prod_{m=1}^{M}\prod_{j=1}^{N}\left(a_{kj}^{(m)}\right)^{\alpha_{m}/N}}{\sum_{i=1}^{N}\prod_{m=1}^{M}\prod_{j=1}^{N}\left(a_{ij}^{(m)}\right)^{\alpha_{m}/N}},
\end{equation}
$k=1,2,\ldots,N$, such that $\sum_{k=1}^{N}w_{k}=1$. 

This result can be obtained also using a different approach, based
on the minimization of the generalized Kullback-Leibler divergence
\cite{key-11}. 

Let us assume that we solve the AHP problem separately
for each expert, characterized by the judgment matrix $A^{(m)}=[a_{ij}^{(m)}]_{N\times N}$,
and we obtain the unnormalized priority vectors: $u^{(m)}=[u_{1}^{(m)},u_{2}^{(m)},\ldots,u_{N}^{(m)}]^{T}$,
$m=1,2,\ldots,M$, with the components: \begin{equation}
u_{k}^{(m)}=\prod_{j=1}^{N}\left(a_{kj}^{(m)}\right)^{1/N},\; k=1,2,\ldots,N.\end{equation}
The generalized Kullback-Leibler divergence between the group-priority
vector $u=[u_{1},u_{2},...,u_{N}]^{T}$ and the priority vector $u^{(m)}$
of the expert $m$, is given by:
$$
D(u||u^{(m)})=
$$
\begin{equation}
=\sum_{j=1}^{N}u_{j}\log\left(\frac{u_{j}}{u_{j}^{(m)}}\right)-\sum_{j=1}^{N}u_{j}+\sum_{j=1}^{N}u_{j}^{(m)}.
\end{equation}
The problem is to find the vector $u$, which minimizes the weighted
sum of generalized Kullback-Leibler divergences between $u$ and $u^{(m)}$,
$m=1,2,\ldots,M$: \begin{equation}
D(u_{1},u_{2},\ldots,u_{N})=\sum_{m=1}^{M}\alpha_{m}D(u||u^{(m)}).\end{equation}
The minimum condition is:\begin{equation}
\frac{\partial}{\partial u_{k}}D(u||u^{(m)})=0,\; k=1,2,\ldots,N,\end{equation}
which is equivalent with:\begin{equation}
\log u_{k}-\sum_{m=1}^{M}\alpha_{m}\log u_{k}^{(m)}=0,\; k=1,2,\ldots,N,\end{equation}
and respectively:
$$
u_{k}=\prod_{m=1}^{M}\left(u_{k}^{(m)}\right)^{\alpha_{m}}=
$$
\begin{equation}
=\prod_{m=1}^{M}\prod_{j=1}^{N}\left(a_{kj}^{(m)}\right)^{\alpha_{m}/N},\; k=1,2,\ldots,N.
\end{equation}

Obviously, by normalizing $u_{k}$ we obtain the same group-priorities
$w_{k}$, like those obtained using the LLS approach.

\section{Conclusion}

In this paper we have discussed the LLS approach for the AHP
and group-AHP, which provides an exact and unique  solution for the priority vector of the AHP. Also, we have shown that the LLS approach for group-AHP is
equivalent with the minimization of the weighted sum of generalized
Kullback-Leibler divergences, between the aggregated priority vector
and the priorities of each expert.

\end{document}